\documentclass[prl,twocolumn]{revtex4}
\usepackage{amsthm}
\usepackage{amsmath}
\usepackage{latexsym}
\usepackage{amsfonts}
\usepackage{amssymb}
\usepackage{color}
\usepackage{bbm,dsfont}
\usepackage{graphicx}
\usepackage{hyperref}
\usepackage{subfigure}

\newcommand{\id}{\mathbbm{1}} 

\begin{document}
\title{Quantum Resource Control for noisy EPR-steering with qubit measurements}
\author{Jukka Kiukas}
\author{Daniel Burgarth}
\affiliation{Department of Mathematics, Aberystwyth University, Aberystwyth SY23 3BZ, UK}
\begin{abstract} We demonstrate how quantum optimal control can be used to enhance quantum resources for bipartite one-way protocols, specifically EPR-steering with qubit measurements. Steering is relevant for one-sided device-independent key distribution, the realistic implementations of which necessitate the study of noisy scenarios. So far mainly the case of imperfect detection efficiency has been considered; here we look at the effect of dynamical noise responsible for decoherence and dissipation. In order to set up the optimisation, we map the steering problem into the equivalent joint measurability problem, and employ quantum resource-theoretic robustness monotones from that context. The advantage is that incompatibility (hence steerability) with arbitrary pairs of noisy qubit measurements has been completely characterised through an analytical expression, which can be turned into a computable cost function with exact gradient. Furthermore, dynamical loss of incompatibility has recently been illustrated using these monotones. We demonstrate resource control numerically using a special gradient-based software, showing, in particular, the advantage over naive control with cost function chosen as a fidelity in relation to a specific target. We subsequently illustrate the complexity of the control landscapes with a simplified two-variable scheme. The results contribute to the theoretical understanding of the limitations in realistic implementations of quantum information protocols, also paving way to practical use of the rather abstract quantum resource theories.

\

\end{abstract}

\maketitle

\begin{figure}
\begin{center}
\includegraphics[scale=0.6]{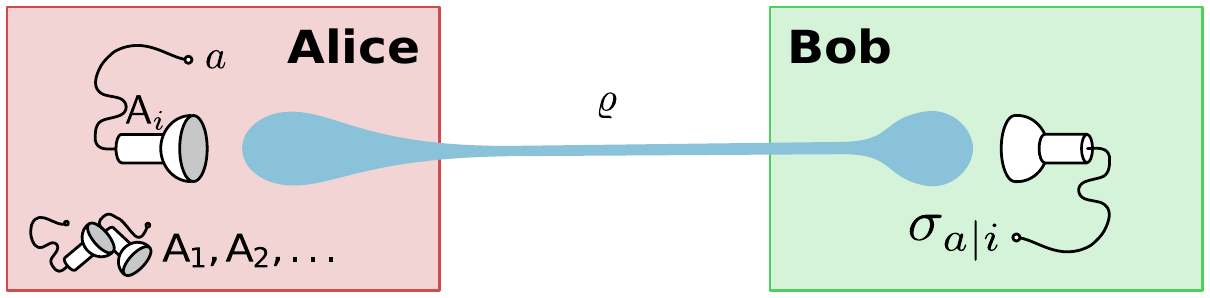}
\end{center}
\caption{\label{fig:steering} A correlation experiment for EPR-steering.}
\end{figure}

\paragraph{Introduction---}Due to the emerging technological motivation, it has become popular to view quantum effects as \emph{resources} for tasks which cannot be described using classical physics \cite{brandao, spekkens, Horo09}. While most work focuses on non-classical properties of quantum states, measurement resources are just as important, since the set of available measurements in any real experiment is restricted by the implementable \emph{controls} such as laser pulses \cite{DoPe10}. This is particularly relevant in correlation experiments where local parties make measurements on a shared entangled state. If the correlations violate a Bell inequality, they can be used in Quantum Key Distribution without any knowledge of the measurement devices; this is however experimentally difficult due to the detection loophole \cite{AcBr07, HeBe15}. Implementation is less challenging in a semi-device-independent scenario based on Einstein-Podolsky-Rosen (EPR) steering \cite{WiJoDo07} (Fig. \ref{fig:steering}), which has attracted considerable interest recently \cite{Sk14, Pu13, KoSk15, BrCa12, PiWa14, ibc, UoMoGu14, QuVeBr14, UoCo15}. It is intriguing as it requires entanglement but can be done with correlations admitting a hidden variable model. For instance, steering is possible with Gaussian states and measurements \cite{WiJoDo07} which cannot violate Bell inequalities due to the hidden variable model provided by the Wigner function.

The intuitive idea is Alice "steering" Bob with her measurements through the shared state, which "transmits" an "assemblage" of conditional states to Bob \cite{WiJoDo07, Sk14}. The maximally entangled state provides perfect transmission, and the general case reduces to that by replacing Alice's measurements by certain state-dependent ones Bob reconstructs from the assemblage \cite{UoCo15}. Hence, the quantum resource for steering can be described entirely by measurements; this is very useful when treating the loss of steerability due to local noise, as we see below. Interestingly, the required measurement resource has an independent meaning \cite{UoMoGu14, QuVeBr14, UoCo15}: the measurements need to be \emph{incompatible}. Here incompatibility does not mean non-commutativity, as noisy measurements are typically not projective. It means the non-existence of a "hidden" measurement jointly simulating all Alice's measurements; this notion has been studied for a long time \cite{Lu64, LaPu97, BuLaMi96, siq, WoPeFe09, ReReWo13, HeKiRe14, Busch, StReHe08}.

Hence, the loss of steerability can be described independently of the bipartite scenario, as the loss of incompatibility on Bob's side. This leads to a simplification in system size, the simplest case being a single qubit. In order to \emph{quantify} this loss, we need a numerical \emph{incompatibility monotone}; they can be constructed \cite{HeKiRe15,UoCo15} using the noise-robustness idea from general quantum resource theories \cite{brandao, GrPo05, AlPi07, Pu15, BuHe13, Ha15}. The dynamics of incompatibility has recently been studied using these monotones \cite{AdHe15}.

In this paper, we take a new direction by showing how steering resource can be \emph{directly} enhanced in the presence of Markovian noise, using numerical gradient-based quantum optimal control \cite{DoPe10, RaVi00, KhReKe05}. Research in this area aims at characterising operations reachable with restricted set of controls such as laser pulses, and numerically finding optimal pulses implementing a given target. While unitary control is fairly well established, control of noisy operations (quantum channels) is more challenging due to their complicated structure even in small systems \cite{DiBuMi15, OmDi12, ArGu14, ScSp11}. In contrast to the usual optimisation of a distance from a specific target, we optimize over an incompatibility monotone, so as to do \emph{purpose-oriented control} of EPR-steering, in analogy to entanglement control \cite{KrCi01,PlMi10, MaWi07}. Steering is more challenging in small systems, where the existence of a suitable hidden variable model is already a nontrivial question; we use the special characterisation of qubit incompatibility \cite{Busch, StReHe08} to compute a faithful incompatibility monotone with exact gradient. The purpose-oriented control of steering rather than targeting specific measurements is motivated by the fact that \emph{many} measurements have equal steering potential, and that targeting specific ideal (projective) ones is not likely to work as they are usually not reachable in noisy systems. The problem is also intriguing in that the monotones are unitary-invariant; unitary control \emph{can only help in the presence of noise} which destroys the resource in the first place; hence it is a priori not at all clear if it actually does help.

\paragraph{The quantum resource for steering---}We look at the bipartite scenario with Alice and Bob sharing a state $\rho$ on the tensor product Hilbert space $\mathcal H_A\otimes \mathcal H_B$. Alice has a restricted set $\mathcal M_A=\{ \mathsf A_1,\ldots, \mathsf A_n\}$ of measurements, assumed to be general (possibly non-projective) POVMs, as this is necessary in noisy scenarios. Hence each $\mathsf A_i$ stands for a collection of positive-semidefinite matrices $A_i(a)$ with $\sum_a A_i(a)=\id$. The steering protocol proceeds as follows \cite{WiJoDo07}: Alice chooses index $i$, performs $\mathsf A_i$ on her system getting an outcome $a$, and announces $(i,a)$ to Bob, who uses state tomography to extract the "assemblage" \cite{Pu13} $\sigma_{a|i}:={\rm tr}_A[\rho (A_i(a)\otimes \id)]$ of conditional states. The assemblage is called steerable if Bob cannot reconstruct it from some pre-existing collection of "hidden" states using only classical information on Alice's results. We state the precise definition in terms of the associated joint measurability problem \cite{UoCo15,UoMoGu14, QuVeBr14}: following \cite{UoCo15} we restrict w.l.o.g. $\mathcal H_B={\rm ran} \rho_B$, and define
$$
S_{\rho}(A) := \rho_B^{-\frac 12}{\rm tr}_A[\rho (A\otimes \id)] \rho_B^{-\frac 12},
$$
so that $S_\rho$ is a unital CP map between Alice's and Bob's observable algebras, and ${\rm tr}[\rho (\id\otimes S_{\rho}(A^\intercal))] = {\rm tr}[\rho(A\otimes \id)]$ for any matrix $A$, where $A^\intercal$ is the transpose. This means that Bob can simulate Alice's measurements via the POVMs
$\mathsf B_i^\rho = S_{\rho}(\mathsf A_i^\intercal)$ determined by the assemblage. The setting is \emph{non-steerable} if these are \emph{jointly measurable} in that each $\mathsf B_i^\rho$ is a classical probabilistic postprocessing of a single POVM $G(\lambda)$; formally,
$$
\mathsf B_i^\rho(a) = \sum_\lambda p(a|i,\lambda) G(\lambda), \quad \text{ for all }i=1,\ldots,n,
$$
where $\sum_a p(a|i,\lambda)=1$. This definition is equivalent to the usual notion of joint measurability via marginals \cite{ibc}.

In conclusion, the quantum resource for EPR-steering can be characterised as the opposite of joint measurability, often called \emph{incompatibility}, of the collection
$$\mathsf B_i^\rho = S_{\rho}(\mathsf A_i), \quad i=1,\ldots, n$$
of measurements. This formulation has the advantage of referring only to a single system; the "nonlocal" aspect is encapsulated in $S_\rho$. This is especially useful in describing local noise: given a channel $\Lambda$ on Alice's side changing the state as $\rho\mapsto \tilde \rho =(\Lambda\otimes {\rm Id})(\rho)$, the assemblage changes
$$\sigma_{a|i}\mapsto \tilde \sigma_{a|i} = {\rm tr}_A[\rho (\Lambda^*(A_i(a))\otimes \id)].$$
Then the resource transforms into
\begin{equation}\label{dynmeas}
\tilde{\mathsf B}_i^\rho = S_{\rho}\circ \Lambda^*(\mathsf A_i), \quad i=1,\ldots, n,
\end{equation}
the Heisenberg channel $\Lambda^*$ simply concatenating with $S_\rho$. This has a clear interpretation: $S_{\rho}$ describes \emph{preparation noise} in an imperfect production of a maximally entangled state (for which $S_{\rho}={\rm Id}$), while $\Lambda$ is the subsequent \emph{dynamical noise}. Steerability of the noisy assemblage is equivalent to the incompatibility of \eqref{dynmeas}.

Resource theories also contain the idea of \emph{quantification} \cite{brandao}. Incompatibility of a collection $(\mathsf B_1,\ldots,\mathsf B_n)$ of measurements can be quantified by an \emph{incompatibility monotone} \cite{HeKiRe15}, i.e. a number $\mathcal I(\mathsf B_1,\ldots,\mathsf B_n)$ which is zero exactly in the jointly measurable case, and
\begin{equation}\label{monotonicity}
\mathcal I(\Lambda^*(\mathsf B_1),\ldots,\Lambda^*(\mathsf B_n)) \leq \mathcal I(\mathsf B_1,\ldots,\mathsf B_n)
\end{equation}
for any positive linear map $\Lambda$. This fits well with the steering resource \eqref{dynmeas}, where the total channel $S_{\rho}\circ \Lambda^*$ then effects a \emph{quantitative} loss of the resource. In particular, loss due to continuous dynamics $t\mapsto \Lambda_t$ is described by the function $t\mapsto \mathcal I\left(S_{\rho}\circ \Lambda^*_t(\mathsf A_1), \ldots, S_{\rho}\circ \Lambda^*_t(\mathsf A_n)\right)$.

Good operational incompatibility monotones describe convex-geometric \emph{noise-robustness} \cite{brandao, HeKiRe15,Ha15,UoCo15}: we mix \emph{classical} noise with distribution $p=(p_i)$ into measurements via $\mathsf A\mapsto (1-\lambda)\mathsf A+ \lambda p \id$, and define $\mathcal I$ to be the minimal $\lambda$ for which the resource is lost, i.e. measurements become jointly measurable (see \cite{HeKiRe15,AdHe15} for discussion).

In this paper, we look at the simplest setting with $\mathcal H_A=\mathbb C^2$ and two measurements for Alice; this case is already interesting. The speciality is that robustness monotones can be computed using the analytical characterisation of incompatibility \cite{Busch, StReHe08}: Given a POVM $\mathsf A =(A, \id-A)$ on $\mathbb C^2$, we identify $A = \tfrac 12 (x^0 \id+{\bf x\cdot \sigma})$ with the 4-vector $x=(x^0,{\bf x})$, and $\id-A$ with $x^\perp:=(2-x^0, -{\bf x})$. The condition $0\leq A\leq \id$ reads $x,x^\perp\in \mathcal F_+$, where $\mathcal F_+=\{ x\mid \langle x|x\rangle\geq 0,\, x^0\geq 0\}$ and $\langle x|y\rangle:=x^0y^0-\sum_{i=1}^3 x^iy^i$ is the Minkowski form. A pair of measurements $x_1$ and $x_2$ is jointly measurable \emph{if and only if} $\mathsf C(x,y)\geq 0$, where
\begin{equation}
\begin{split}
&\mathsf{C}(x_1,x_2):=\  \left[ \langle x_1|x_1\rangle\langle x_1^\perp|x_1^\perp\rangle\langle x_2|x_2\rangle\langle x_2^\perp |x_2^\perp\rangle \right]^{1/2}\nonumber\\
&-\langle x_1|x_1^\perp\rangle\langle x_2|x_2^\perp\rangle 
+ \langle x_1|x_2^\perp\rangle\langle x_1^\perp| x_2\rangle+\langle x_1|x_2\rangle\langle x_1^\perp|x_2^\perp\rangle \, .\label{Cfunctional}
\end{split}
\end{equation}
We use the robustness monotone $\mathcal{I}_b(x,y)$ of \cite{HeKiRe15}: given a probability $p=\tfrac 12 (1+b)$, the above classical noise is $x\mapsto N_{\lambda,b}(x):= ((1-\lambda)x^0+\lambda 2p, (1-\lambda){\bf x})$, and $\mathcal{I}_b(x,y)$ is by definition the unique solution $0\leq \lambda\leq 1/2$ of
\begin{equation}\label{iceq}
\mathsf C\left(N_{\lambda,b}(x_1),\,N_{\lambda,b}(x_2)\right)=0.
\end{equation}
Interestingly, $\mathcal{I}_{0}(x_1,x_2)$ coincides with the maximal violation of the CHSH-Bell inequality with Alice's measurements $(x_1,x_2)$ \cite{HeKiRe15}. This monotone was recently used in studying the loss of incompatibility on open systems \cite{AdHe15}.

\paragraph{Optimal resource control of steering---} Having identified and quantified the resource for steering and described its loss in open systems, it is natural to ask if this loss can be slowed down by control. As discussed in the introduction, our approach is to directly optimise the incompatibility resource \eqref{dynmeas} using the monotones $\mathcal I_b$.

We take Alice's noise to be Markovian: $\Lambda_t=e^{t\mathcal L_0}$ with a drift Lindbladian $\mathcal L_0$. We look at two basic cases, amplitude damping $\mathcal L_0^{\rm AD}(\rho) = \gamma (2\sigma_-\rho\sigma_+ -\{\sigma_+\sigma_-,\rho\})$ (containing decoherence and dissipation), and dephasing in the $\sigma_y$-basis $\mathcal L_0^{\rm DP}(\rho) = \gamma (\sigma_y\rho\sigma_y^\dagger -\rho)$ (only decoherence), with $\gamma=0.1$. We model control (e.g. a laser pulse) by changing $\mathcal L_0$ into
$$
\mathcal L_{c} = \mathcal L_0 -i c [H,\cdot],
$$
where the $H$ is a control Hamiltonian and $c\in \mathbb R$. Applying a sequence ${\bf c}=(c_1,\ldots,c_m)$ of such pulses, each of duration $\Delta t$, the dynamics at time $T=m\Delta t$ is $\Lambda_{\bf c} = e^{ \Delta t \mathcal L_{c_m}}\cdots e^{ \Delta t \mathcal L_{c_1}}$. We consider the setting where Alice aims at steering Bob at time $T$ with a measurement pair $(x_1,x_2)$, given that the initial state was $\rho$. According to the preceding section, the information on the resource is faithfully encoded into the cost function
\begin{equation}\label{fidelity}
f({\bf c})=\mathcal I_b(S_\rho\circ \Lambda^*_{\bf c}(x_1),S_\rho\circ \Lambda^*_{\bf c}(x_2))
\end{equation}
describing the steering robustness. In particular, this function is nonzero if and only if the setting is steerable. We implemented \eqref{fidelity} numerically by solving \eqref{iceq} using standard root-finding which needs only a few iterations as the function is just a combination of polynomials and square root, and changes sign on $[0,1/2]$. Given this value, the gradient of $\mathcal I_b$ can be found \emph{analytically} via implicit differentiation. This fits particularly well with the control software \emph{Qtrl} \cite{qtrl}, which implements the Frechet derivative of ${\bf c}\mapsto \Lambda_{\bf c}$ to be used in computing cost functions; hence we get the gradient of $f({\bf c})$ from the chain rule, and employ optimisation based on exact gradient.

\begin{figure}
\begin{center}
\includegraphics[scale=0.4]{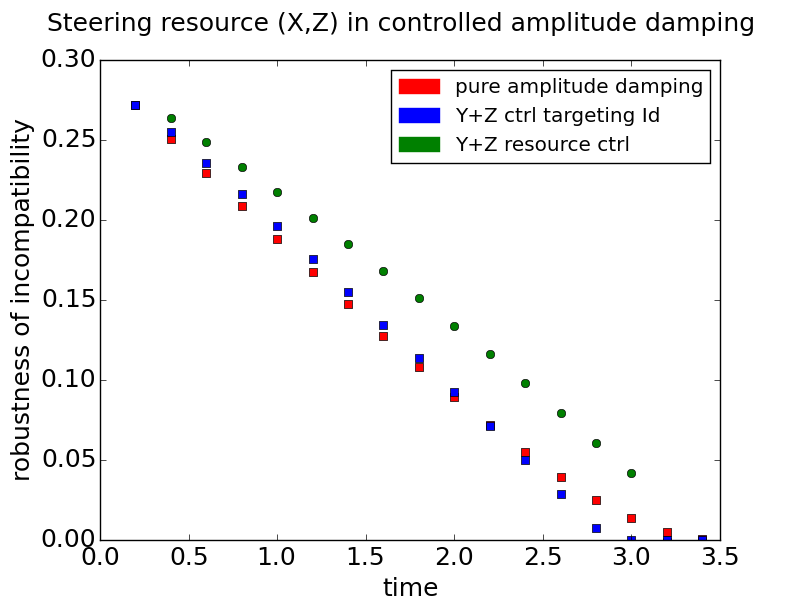}
\end{center}
\begin{center}
\includegraphics[scale=0.4]{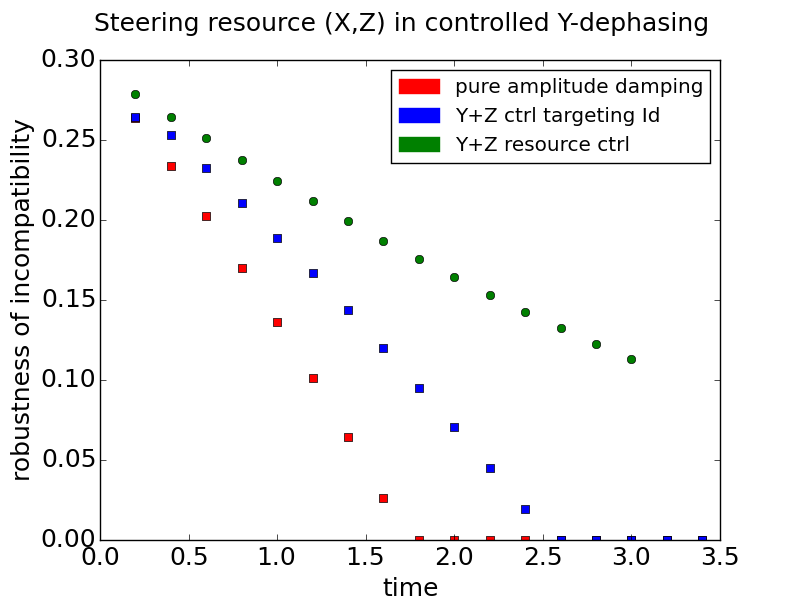}
\end{center}
\caption{\label{fig:ctrlad} Resource control of steering under amplitude damping (upper panel) and dephasing in the $\sigma_y$-basis (lower panel).}
\end{figure}
We take a maximally entangled $\rho$ so that $S_\rho={\rm Id}$, Alice's measurements $(\sigma_x,\sigma_z)$, $H=\sigma_y+\sigma_z$, and use the monotone $\mathcal I_0$. The results for different times $T$ are depicted in Fig. \ref{fig:ctrlad}. It shows the robustness $\mathcal I_0$ in the uncontrolled case, the optimised value, and comparison with the naive control strategy aiming at the channel closest to the identity. The computations were done with \emph{Qtrl} on HPC Wales with $m=20$ (number of time slots) and optimised over $100$ random initial pulses. Optimisation with the cost function \eqref{fidelity} is considerably slower than the naive method, however the results are significantly better. We also see that control works better with dephasing, presumably due to the lack of dissipation present in the amplitude damping.

\begin{figure}
\begin{center}
\includegraphics[scale=0.4]{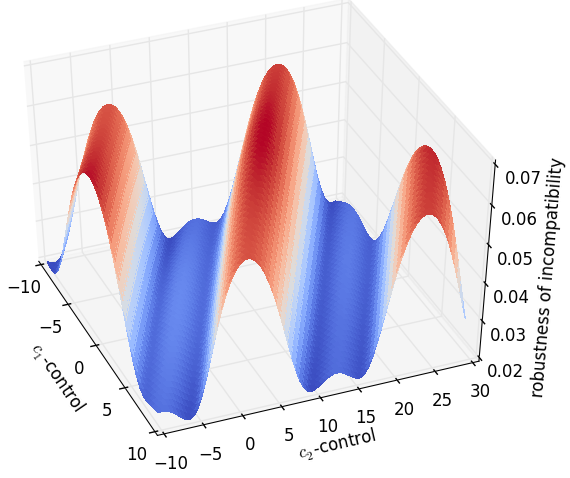}
\end{center}
\caption{\label{fig:landscapeAd} A control landscape of steering robustness under amplitude damping noise. The maximum represents optimal quantum resource control.}
\end{figure}

An inspection of the optimal pulses showed that the amplitudes peak close to the end, suggesting that only a few time slots are needed. Accordingly, we considered the following simple scheme: drift until time $t_{\rm drift}<T$, then apply two pulses $c_1,c_2$, each of duration $\Delta t = (T-t_{\rm drift})/2$. This implements the map $\Lambda_{c_1,c_2}=e^{ \Delta t \mathcal L_{c_2}}e^{\Delta t \mathcal L_{c_1}}e^{t_{\rm drift} \mathcal L_{0}}$. The corresponding control landscape
for $f(c_1,c_2)$ in \eqref{fidelity} is shown in Fig. \ref{fig:landscapeAd} for the amplitude damping case with $t_{\rm drift}=2.6$ and $T=2.8$. At this time resource control can boost steering robustness up to 0.062, which is more than a factor of two improvement from the pure noise (see Fig. \ref{fig:ctrlad}). The maximum is attained with the pulse ${\bf c}=(-1.42, 12.32)$. Clearly the choice of controls is crucial; looking also at Figure \ref{fig:ctrlad} we observe that both pure noise (no control) and the naive strategy are far from the maximum. In the dephasing case the landscape is similar, except that there are controls leading to non-steerability (including the no control and naive control cases), while optimal control manages to preserve steering. The maximum steering robustness is 0.125 with the pulse ${\bf c}=(1.80, -12.88)$. It appears to be crucial that the controls are applied just before measuring, when incompatibility is close to its maximum.

\paragraph{Conclusion and outlook---} We have demonstrated how steering can be enhanced by control in noisy qubit systems, with direct resource optimisation of \eqref{fidelity} performing better than target-based one. The effect of an imperfect initial state $\rho$, and other non-Markovian features \cite{AdHe15} remain to be studied. Our results pave the way for general schemes for implementing optimal noisy quantum resources in controlled open systems. The optimisation naturally becomes slow in large systems, with analytical gradient no longer available. Nevertheless, \eqref{fidelity} can always be computed efficiently via a semidefinite program \cite{HeKiRe15}, and approximations based on steering inequalities \cite{KoSk15} could be used in analogy to the entanglement control \cite{PlMi10} to make the computations feasible.

\paragraph{Acknowledgment---} We thank the QuTiP project \cite{qtrl} for providing the software, (especially A. Pitchford for developing the pulse optimisation package), and HPC Wales for computing time. We acknowledge financial support from the EPSRC project EP/M01634X/1.


\end{document}